\documentclass[12pt,preprint]{aastex}
\usepackage{graphicx}
\usepackage{longtable}
\usepackage{multirow}
\usepackage{url}
\usepackage{lipsum}

\begin{document}

\title{New Model-independent Method to Test the Curvature of Universe}

\author{H. Yu\altaffilmark{1} and F. Y. Wang\altaffilmark{1,2}{*}}
\affil{
$^1$ School of Astronomy and Space Science, Nanjing University, Nanjing 210093, China\\
$^2$ Key Laboratory of Modern Astronomy and Astrophysics (Nanjing University), Ministry of Education, Nanjing 210093, China\\
} \email{{*}fayinwang@nju.edu.cn}

\begin{abstract}
We propose a new model-independent method to test the cosmic
curvature by comparing the proper distance and transverse comoving
distance. Using the measurements of Hubble parameter $H(z)$ and
angular diameter distance $d_A$, the cosmic curvature parameter
$\Omega_K$ is constrained to be $-0.09\pm0.19$, which is consistent
with a flat universe. We also use Monte Carlo simulation to test the
validity and efficiency, and find that our method can give a
reliable and efficient constraint on cosmic curvature. Compared with
other model-independent methods testing the cosmic curvature, our
method can avoid some drawbacks and give a better constraint.
\end{abstract}


\section{Introduction}\label{sec:intro}
The cosmic curvature is a fundamental parameter
for cosmology. Whether the space of our Universe is open, flat, or
closed is important for us to understand the evolution of our
universe, and dark energy equation of state
\citep{Clarkson2007,Zhao2007,Ichikawa2006a,Weinberg2013}. Due to the
strong degeneracy between the curvature and the dark energy equation
of state, it is difficult to study a non-flat $\omega$CDM model
\citep{Clarkson2007}. Besides, a significant detection of a non-zero
curvature will affect the fundamental theory of cosmology because
most observations support a flat $\Lambda$CDM model, including the
latest Planck result which gives $|\Omega_k|<0.005$
\citep{Planck2015}. However, most of these constraints are not in a
direct geometric way. Therefore, determining the cosmic curvature
with model-independent methods is very important.

In order to constrain the cosmic curvature in a direct geometric
way, some definitions of cosmological distance should be introduced.
Several distance definitions, such as the proper distance $d_P$,
luminosity distance $d_L$, angular diameter distance $d_A$ and
transverse comoving distance $d_M$ are defined to investigate
cosmology \citep{Hogg1999,Coles2002,Weinberg2008,Weinberg2013}. Under
the assumption of Friedmann-Lema\^{\i}tre-Robertson-Walker metric,
the proper distance can be expressed as
\begin{equation}\label{dP1}
    d_P(r) = a_0\int^r_0\frac{dr^\prime}{\sqrt{1-Kr^{\prime2}}}=a_0f(r),
\end{equation}
with $f(r)=\sin^{-1}r$, $r$, or $\sinh^{-1}r$ for curvature $K=+1$,
$0$, or $-1$, $a_0$ is the present scale factor, and $r$ is the
comoving coordinate of the source. With the definition of Hubble
parameter $H(z)=\dot{a}/{a}$, it can also be expressed as
\begin{equation}\label{dP2}
    d_P(z) = \frac{c}{H_0}\int^z_0\frac{dz^\prime}{E(z^\prime)},
\end{equation}
where $z$ is the redshift, $H_0$ is the Hubble constant, $c$ is the
speed of light and $E(z)=H(z)/H_0$. Similarly, the transverse
comoving distance can be expressed as
\begin{equation}\label{dM}
    d_M(z)=a_0r(z)=\frac{c}{H_0\sqrt{-\Omega_K}}
    \sin[\sqrt{-\Omega_K}\int^z_0\frac{dz^\prime}{E(z^\prime)}],
\end{equation}
where $\Omega_K$ is the energy density of cosmic curvature
($-i\sin(ix)=\sinh(x)$ if $\Omega_K>0$). With the definition of
$d_M$, $d_L$ and $d_A$ can be derived through $d_L=d_M(1+z)$ and
$d_A=d_M/(1+z)$, respectively.

Numerous works have been done to determine the curvature parameter
$\Omega_K$ using different methods, some of which are
model-independent. \cite{Bernstein2006} proposed a model-independent
method using the weak lensing and baryon acoustic oscillation (BAO)
data to constrain $\Omega_K$ based on the distance sum rule, which
was used to test the FLRW metric in \cite{Rasanen2015} (hereafter,
called DSR method). The basic principle of DSR method is that the
relation between $d(z_s)$ and $d(z_l)+d(z_l,z_s)$ depends on the
cosmic geometry (see Fig. 1 of \cite{Bernstein2006}). The value of
$d(z_l,z_s)$ can be calculated from gravitational lensing. However,
the large uncertainty in gravitational lens system restricts its
efficiency on constraining the curvature \citep{Rasanen2015}. Another
important model-independent method was proposed in
\cite{Clarkson2007}, by comparing the Hubble parameter $H(z)$ and
the derivative function of transverse comoving distance $d_M$ gained
from $d_A$, which has been used in many works
\citep{Clarkson2008,Yahya2014,Li2014,Cai2016} (hereafter, C07
method). The basis of this method is that one can determine the
curvature by combining measurements of the Hubble parameter $H(z)$
and the transverse comoving distance $d_M(z)$
\begin{equation}\label{C07}
    \Omega_K=\frac{[H(z)d_M^\prime(z)]^2-c^2}{[H_0d_M(z)]^2},
\end{equation}
where $^\prime$ means the derivative with respect to redshift $z$.
However, in this method, one needs to determine the derivative
function of transverse comoving distance $d_M$ from a fitting
function, which will introduce a large uncertainty.

Therefore, in order to avoid the drawbacks of the two methods, we
propose a new direct geometric method to test the cosmic curvature.
This method is based on the comparison between proper distance $d_P$
obtained from Hubble parameter measurement, and transverse comoving
distance $d_M$ obtained from angular diameter distance $d_A$
measurement. The structure of this paper is organized as follows. In section
\ref{sec:method_omega_k}, we introduce our new model-independent
method to test the cosmic curvature using $d_P$ and $d_M$. In
section \ref{sec:result}, we give our constraint on $\Omega_K$ using
the Hubble parameter and angular diameter distance measurements. In
section \ref{sec:simulation}, we test the validity and efficiency of
our method with Monte Carlo simulation. In section
\ref{sec:comparison}, we discuss its advantages compared with other
methods. Finally a summary will be given in section
\ref{sec:summary}.

\section{Method to test $\Omega_K$}\label{sec:method_omega_k}
Comparing the definitions of proper distance $d_P$ and transverse
comoving distance $d_M$, one can find that the difference between
them only caused by the curvature of universe. This gives the basis
to test the cosmic curvature using the comparison of $d_P$ and
$d_M$. From equations (\ref{dP2}) and (\ref{dM}), $\Omega_K$ can be
derived from
\begin{equation}\label{dpdm}
\frac{H_0d_M}{c}\sqrt{-\Omega_K} = \sin(\frac{H_0d_P}{c}\sqrt{-\Omega_K}).
\end{equation}
Equation (\ref{dpdm}) gives the direct relation among $\Omega_K$,
$d_P$ and $d_M$. Once the $d_P$ and $d_M$ are determined, $\Omega_K$
can be calculated through this equation. If the value of redshift
and $\Omega_K$ are not large, $H_0d_P\sqrt{-\Omega_K}/c$ is less
than one. From the Taylor expansion, the equation (\ref{dpdm}) can
be approximated as
\begin{equation}\label{omega_k}
\Omega_K = \frac{6c^2}{H_0^2}\frac{d_M-d_P}{d_P^3},
\end{equation}
from which $\Omega_K$ can be determined directly.

\begin{figure}
  \includegraphics[width=\textwidth]{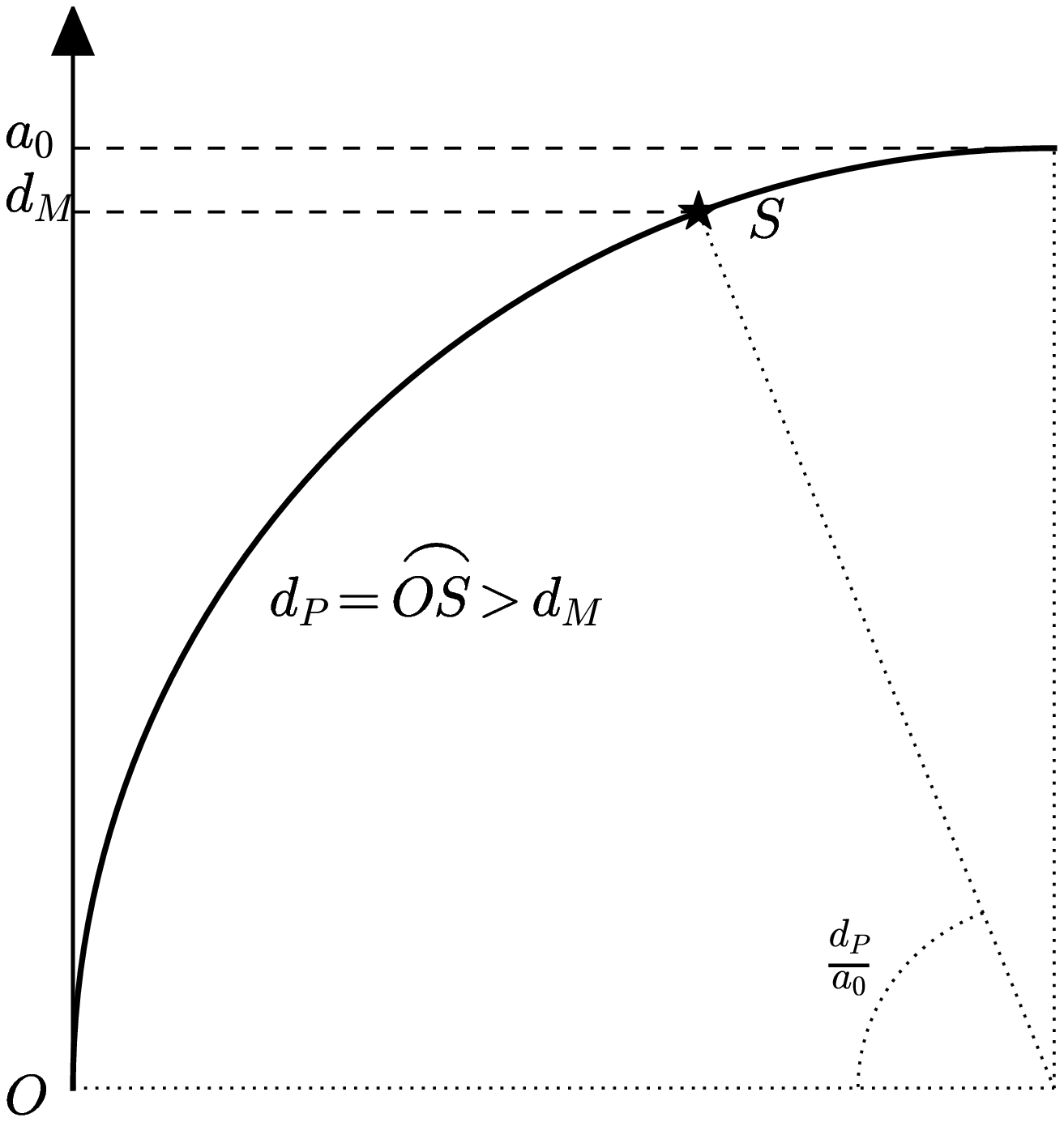}\\
  \caption{Illustration of the proper distance $d_P$ and transverse comoving distance $d_M$
  in a closed universe. It is obvious that $d_M<d_P$.}\label{fig1}
\end{figure}

Figure \ref{fig1} shows the key principle of our method in the
$\Omega_K<0$ case. In this figure, the arc OS is the proper distance
between source $S$ and observer $O$, while the transverse comoving
distance between them is $d_M=a_0\sin(\frac{d_P}{a_0})$. In this
case, it is obvious that the $d_M$ of an object has an up limit
$a_0$ and it is less than $d_P$. In contrast, in an open universe
($\Omega_K>0$), we have $d_M>d_P$. $d_M=d_P$ only happens in a flat
universe. Therefore, the cosmic curvature can be derived by
comparing $d_M$ and $d_P$. Undering the assumption of
$\sqrt{|\Omega_K|}I\ll1$, we can obtain
\begin{equation}\label{delta}
    \delta\equiv \frac{d_P-d_M}{d_P} = -\frac{\Omega_KI^2}{6},
\end{equation}
with $I=\int^z_0\frac{dz^\prime}{E(z)}$. The $\delta$ means the
relative difference between $d_P$ and $d_M$. The value of $|\delta|$
gives the requirement on the accuracy of measurement, which means
the $\Omega_K$ cannot be constrained if the observed uncertainty is
much larger than $|\delta|$. Besides, if the total relative error of
the measurement sample of $d_M$ and $d_P$ is $\sigma$, one can
expect that the tightest constrain on $\Omega_K$ will have an error
about $\sigma_{\Omega_K}\sim6\sigma/I^2$. In other words, equation
(\ref{delta}) gives the constraint limit of this method.

To obtain the transverse distance $d_M$, we choose the angular
diameter distance $d_A$ measurement based on BAO in several previous
works \citep{Blake2012,Xu2013,Samushia2014,Delubac2015}. These data
and their references are listed in Table 1. The detailed information
about these data can be found in their references. The $d_M$ can be
easily derived with the direct relation between them $d_M=d_A(1+z)$.
The next important issue is how to measure $d_P$. From equation
(\ref{dP2}), the proper distance $d_P$ only depends on the $H(z)$
function. Therefore, in order to derive the proper distance $d_P$,
one can construct the $H(z)$ function from Hubble parameter
measurements. Then $d_P$ can be derived from equation (\ref{dP2}).
There are tens of Hubble parameter measurements derived from
differential ages of galaxies and the radial BAO in the previous
literature, which are listed in Table 2. In order to make our method
model-independent, Gaussian Process (GP) method is used to
reconstruct the $H(z)$ function. GP method is a powerful tool to
reconstruct a function from data directly without any assumption of
the function form and is used widely in astronomy
\citep{Holsclaw2010,Shafieloo2010,Shafieloo2012,Seikel2012a,Bilicki2012}.
Therefore, with GP method, we don't need any prior cosmological
model. There is a good python package for GP method called Gapp
developed by \cite{Seikel2012a} which was used in many works
\citep{Seikel2012b,Bilicki2012,Cai2016}. It can reconstruct the
function as long as observed data was input. More detailed
information about GP method and Gapp can be found in
\cite{Seikel2012a}.

The main route of our method is that: I) deriving the transverse
comoving distance $d_M$ from angular diameter distance $d_A$
measurements; II) reconstructing $H(z)$ function from Hubble
parameter measurements using GP method; III) using equation
(\ref{dP2}) to calculate the proper distance at a certain redshift;
IV) using equation (\ref{dpdm}) or (\ref{omega_k}) to determine
$\Omega_K$ at a certain redshift. Several $\Omega_K$ can be
determined at different redshifts since there are several $d_A$
measurements. One can choose the average of them as the final result
through
\begin{equation}\label{ave_omega_K}
    \Omega_K = \frac{\sum_i\limits\Omega_{K,i}/\sigma^2_{\Omega_{K,i}}}{\sum_i\limits1/\sigma^2_{\Omega_{K,i}}}
\end{equation}
where $\Omega_{K,i}$ and $\sigma_{\Omega_{K,i}}$ are determined
$\Omega_K$ and its uncertainty at a certain redshift. The total
uncertainty can be obtained from
\begin{equation}\label{ave_sigma_omega_K}
    \sigma_{\Omega_K}^2 = \frac{1}{\sum_i\limits1/\sigma^2_{\Omega_{K,i}}}.
\end{equation}

\section{Results}\label{sec:result}
\begin{figure}
  \includegraphics[width=\textwidth]{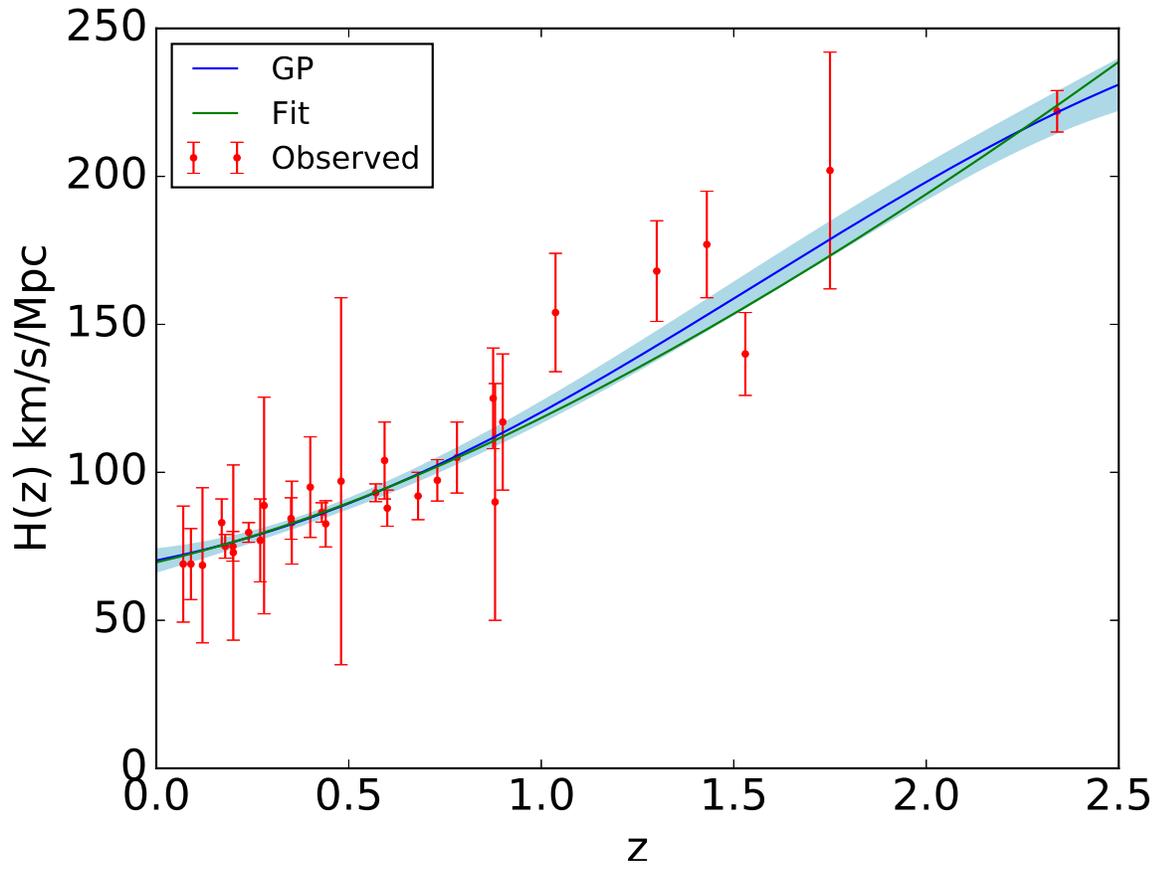}\\
  \caption{The blue curve and area show the $H(z)$ function and its $1\sigma$ confidence region re-constructed from GP method.
  The green curve shows the $H(z)$ function fitted with $\Lambda$CDM model. The red points and the error bars show the observed Hubble parameters and their $1\sigma$ errors.}\label{fig2}
\end{figure}

\begin{figure}
  \includegraphics[width=\textwidth]{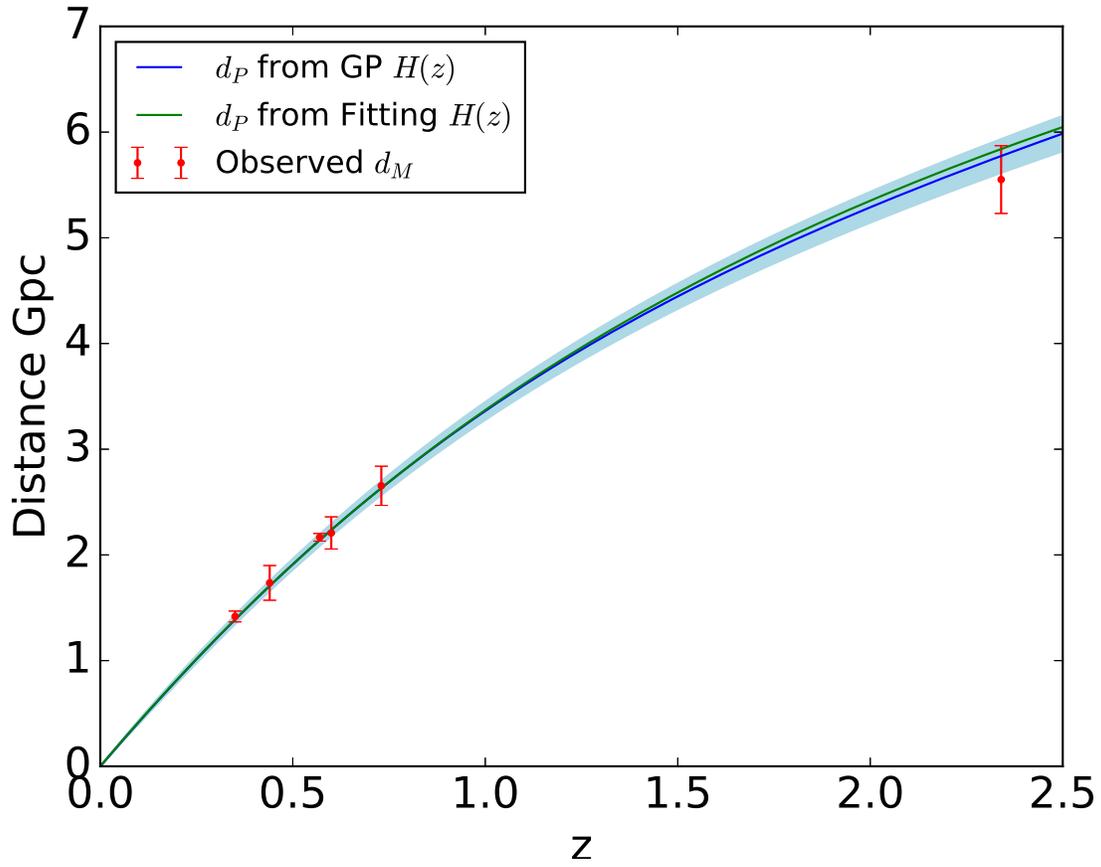}\\
  \caption{The blue curve and area show the $d_P(z)$ function and its $1\sigma$ confidence region derived from the GP-$H(z)$ function.
  The green curve shows the $d_P(z)$ function derived from the fitted $H(z)$ function. The red points and the error bars show the observed $d_M$ and their $1\sigma$ errors.}\label{fig3}
\end{figure}

We collect 31 Hubble parameter measurements from previous literature
and list them in table 2. These Hubble parameters at different
redshifts are derived using differential ages of galaxies and the
radial BAO method. Using these observed data and GP method, the
$H(z)$ function can be reconstructed, which is shown as the blue
curve in Figure \ref{fig2}. Hereafter, this $H(z)$ function is
called GP-$H(z)$. For comparison, we also fit the observed data
based on $\Lambda$CDM model which is shown as the green curve in
Figure \ref{fig2}. Hereafter, this $H(z)$ function is called
Fit-$H(z)$. We can find that the Fit-$H(z)$  is well covered by the
GP-$H(z)$ function and its $1\sigma$ confidence region. With the
reconstructed GP-$H(z)$ function, one can use equation (\ref{dP2})
to derive the proper distance at a certain redshift. The derived
$d_P(z)$ functions from GP-$H(z)$ and Fit-$H(z)$ are shown in Figure
\ref{fig3}. Hereafter, they are called GP-$d_P(z)$ and Fit-$d_P(z)$
respectively. Comparing the GP-$H(z)$ and Fit-$H(z)$, one can find
that the derivation between them becomes smaller than that between
GP-$H(z)$ and Fit-$H(z)$.

\begin{figure}
  \includegraphics[width=\textwidth]{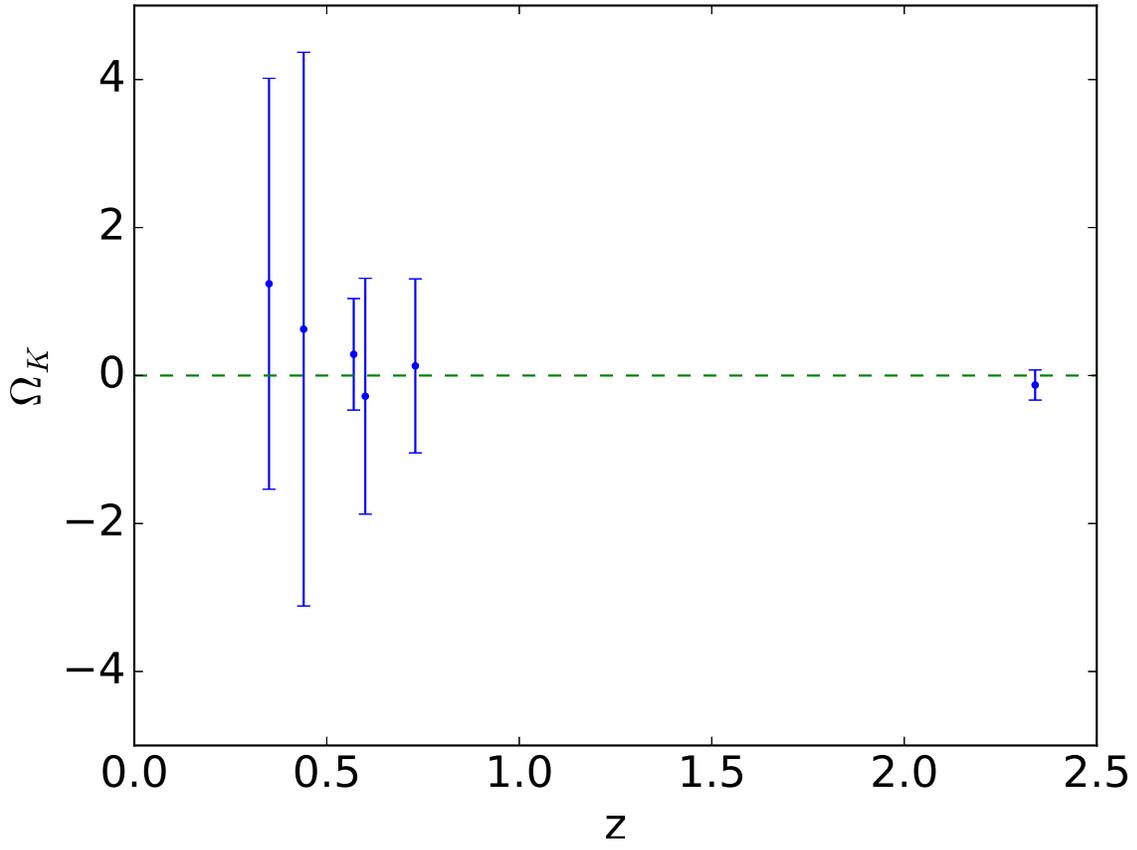}\\
  \caption{The $\Omega_K$ determined by comparing the GP-$d_P(z)$ function and observed $d_M$.}\label{fig4}
\end{figure}

Using equation (\ref{dpdm}), one can derive the $\Omega_K$ from the
$d_P$ and $d_M$ at the same redshift. Figure \ref{fig4} shows the result of derived $\Omega_K$ at different
redshifts. The average $\Omega_K$ and its error bar are derived from
equations (\ref{ave_omega_K}) and (\ref{ave_sigma_omega_K}). The
result is listed in Table 3. The average $\Omega_K$ constrained by
these six $d_A$ data is $\Omega_K=-0.09\pm0.19$. There is no
significant deviation from a flat universe. From bottom panel of
Figure \ref{fig2}, it is obvious that the high-redshift measurement
gives tighter constraint on $\Omega_K$ than the low-redshift
measurement. The reason is that the $I$ term in equation
(\ref{delta}) is larger at high redshift, which will decrease the
error of $\Omega_K$ through $\sigma_{\Omega_K}\sim6\sigma/I^2$.

\section{Simulation}\label{sec:simulation}
In order to test the validity and efficiency of
our method, we perform Monte Carlo simulation. The route of
simulation is as follows: I) creating mock $H(z)-z$ and $d_M-z$ data
sets based on a prior cosmological model; II) reconstructing the
GP-$H(z)$ function from GP method; III) using GP-$H(z)$ function to
derive the GP-$d_P(z)$ function through equation (\ref{dP2}); IV)
using equations (\ref{dpdm}) and (\ref{ave_omega_K}) to constrain
$\Omega_K$ and its average value; V) simulating $10^4$ times for
each prior cosmological model and give the distribution of
determined average $\Omega_K$.

For simualtion, we choose $\Lambda$CDM model as the prior
cosmological model. The model parameters are chosen as $H_0=70~\rm
km/s/Mpc$, $\Omega_M=0.3$ and $\Omega_\Lambda=1-\Omega_M-\Omega_K$
where $\Omega_K=-0.1$, $0$ and $0.1$ for different cases. In each
simulation, there are 20 mock $H-z$ and $d_M-z$ data sets
respectively. The redshifts of these mock data are chosen equally in
$\log(1+z)$ space in redshift range $0.1\le z\le5.0$. The relative
uncertainty of these mock data is $1\%$ which will be realized in
future observation \citep{Weinberg2013}.

Figure \ref{fig5} gives an example of the simulations in the
$\Omega_K=0$ case. The three panels show the mock Hubble parameter
data with GP-$H(z)$ function, mock $d_M$ data with GP-$d_P(z)$
function and the final determined $\Omega_K$. From this figure, we
can see that GP method can reconstruct $H(z)$ function well. In this
case, the final derived average $\Omega_K$ is
$\Omega_K=0.0001\pm0.0092$. Figure \ref{fig6} shows the posterior
distributions of $\Omega_K$ for three $\Omega_K$ cases. From this
figure, one can find that our method can give a reliable and tight
constraint on the prior $\Omega_K$. The uncertainty is
$\sigma_{\Omega_K}\approx0.011$. This result means that if there are
20 $d_A$ and $H(z)$ measurements with $1\%$ uncertainty, our method
can give a constraint on $\Omega_K$ at $1\%$ level. In future, there
will be more accurate measurements of $d_A$ and $H(z)$, tighter
constraint on $\Omega_K$ can be expected.

\section{Compared with other methods}\label{sec:comparison}
In this section, we compare
our method with other model-independent methods. Just as introduced
in the first section, there are two model-independent methods to
constrain the curvature of universe proposed in previous literature
\citep{Bernstein2006,Clarkson2007}, which have been used in many
works \citep{Clarkson2008,Yahya2014,Li2014,Rasanen2015,Cai2016}. The
first one is DSR method based on the distance sum rule proposed by
\cite{Bernstein2006} and the other is C07 method based on the
equation (\ref{C07}) \citep{Clarkson2007}.

The basic principle of DSR method is the distance sum rule which
means that if there are two sources $S_1$ and $S_2$ at redshift
$z_1$ and $z_2$ (assuming $z_1<z_2$), it has
$d(z_2)=d(z_1)+d(z_1,z_2)$ if our universe is flat. Otherwise,
$d(z_2)>d(z_1)+d(z_1,z_2)$ or $d(z_2)<d(z_1)+d(z_1,z_2)$ for
$\Omega_K>0$ or $\Omega_K<0$ respectively. The distance between
$S_1$ and $S_2$ can be determined by gravitational lens
\citep{Bernstein2006,Rasanen2015}. Compared the Figure 1 of
\cite{Bernstein2006} with our Figure 1, it can be found that the
difference between $d_P$ and $d_M$ is larger than that between
$d(z_1)+d(z_1,z_2)$ and $d(z_2)$, which means our method is more
sensitive on the cosmic curvature. An accurate calculation gives the
relative difference between $d(z_1)+d(z_1,z_2)$ and $d(z_2)$ is
\begin{equation}\label{delta_C07}
    |\delta_{DSR}| \le |\frac{\Omega_KI^2}{8}|,
\end{equation}
where $=$ is only valid  when the source $S_1$ locates at the middle
of $S_2$ and observer. Compared with equation (\ref{delta}), one can
find that DSR method needs a higher accuracy of measurement than our
method. Besides, the systematic uncertainty of gravitation lens
system parameter $f^2$ is about $20\%$ \citep{Kochanek2000}, where
$f$ is a phenomenological coefficient that parameterizes uncertainty
due to difference between the velocity dispersion of the observed
stars and the underlying dark matter, and other systematic effects
in strong lensing. This systematic uncertainty is hardly to remove,
which restricts its efficiency on constraining the cosmic curvature.

\begin{figure}
  \includegraphics[width=\textwidth]{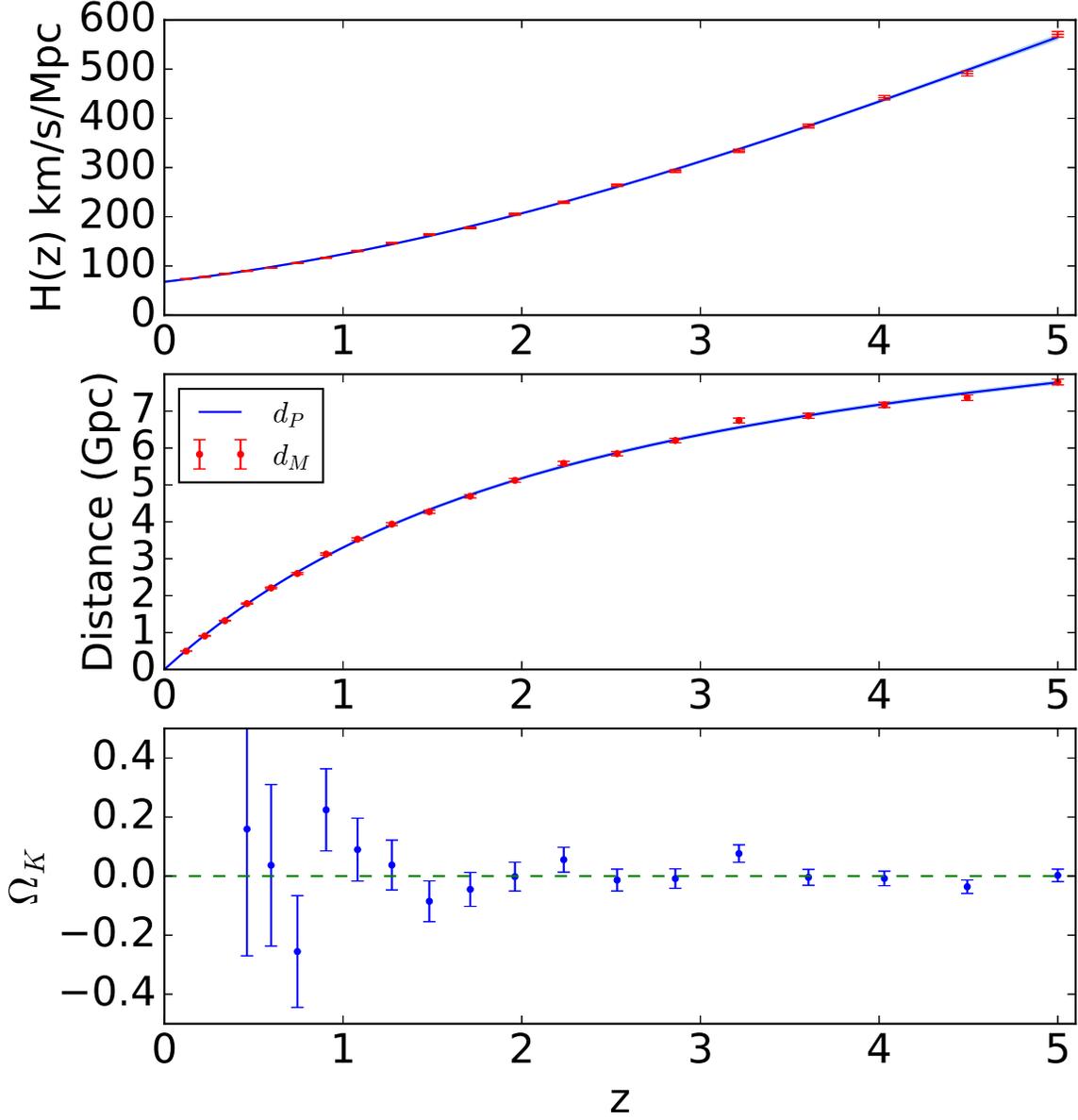}\\
  \caption{An example of the simulation for $\Omega_K=0$ case. Top panel shows the mock Hubble parameter data and the GP-$H(z)$ function. Middle panel shows the mock $d_M$ data and the GP-$d_P(z)$ function.
  The bottom panel shows the final $\Omega_K$ determined from these mock data.}\label{fig5}
\end{figure}

\begin{figure}
  \includegraphics[width=\textwidth]{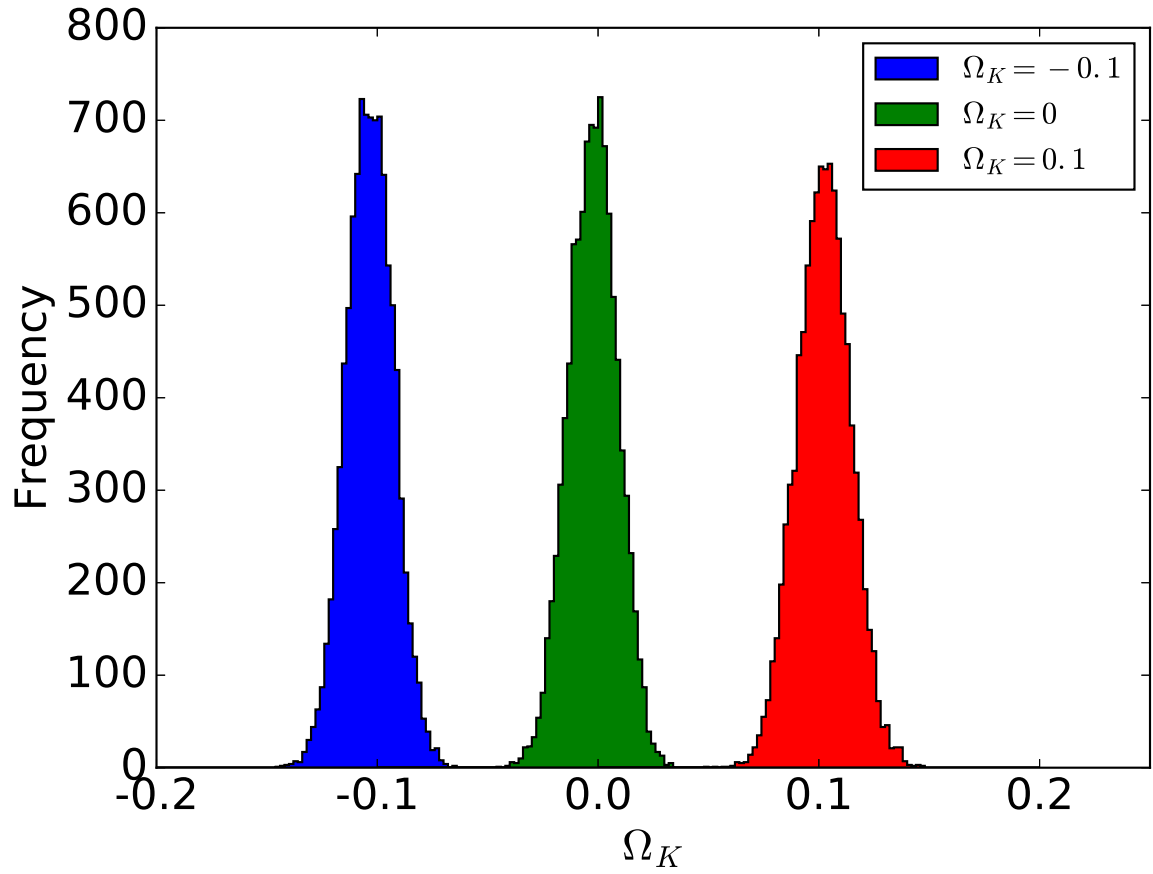}\\
  \caption{The distributions of $\Omega_K$ determined from simulated mock data based on background $\Lambda$CDM model with different prior curvatures using our method.}\label{fig6}
\end{figure}

\begin{figure}
  \includegraphics[width=\textwidth]{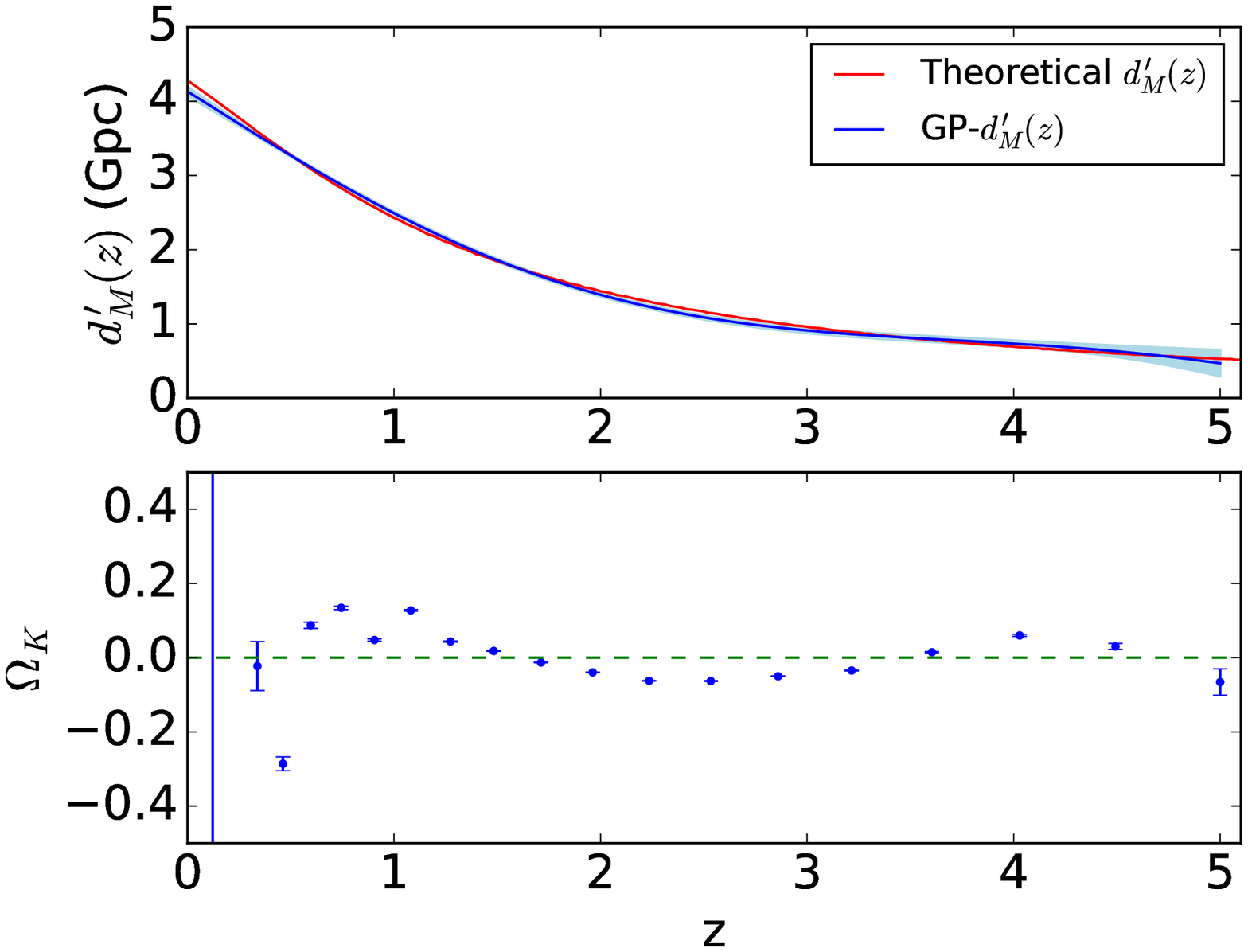}\\
  \caption{Same as Fig. \ref{fig5} but for C07 method.}\label{fig7}
\end{figure}

For the C07 method, it is based on equation (\ref{C07}), which needs
the first derivative function of $d_M(z)$ fitted from observational data.
As is known to all that calculating the derivative function from a
fitted function will increase the uncertainty significantly
especially when the function form is unknown and the data is not
enough. In order to check the efficiency of C07 method, we also use
Monte Carlo simulation to test it, and the simulation is same as
introduced in section \ref{sec:simulation}. Figure \ref{fig7} gives
an example of the simulations in the $\Omega_K=0$ case, which is
similar with Figure \ref{fig5}. Instead of $H(z)$ and $d_P(z)$
functions, we show the $d_M^\prime(z)$ function in the top panel of
Figure \ref{fig7}. Because the $H(z)$ and $d_P(z)$ functions are
similar with those in Figure \ref{fig5}. From Figure \ref{fig7}, it
is obvious that the $d_M^\prime(z)$ function derived from GP method
has a large derivation with the theoretical one, and the determined
$\Omega_K$ are not as well as those in bottom panel of Figure
\ref{fig5}. Figure \ref{fig8} shows the posterior distributions of
$\Omega_K$ determined with C07 method for three $\Omega_K$ cases.
From this figure, one can find that C07 method gives a large
uncertainty on the determined $\Omega_K$.

\section{Summary}\label{sec:summary}
We have proposed a new model-independent method to test the cosmic
curvature in this paper. The main principle of our method is to
compare the proper distance $d_P$ and transverse comoving distance
$d_M$ at same redshift (Figure \ref{fig1} gives an illustration).
Using equation (\ref{dpdm}), one can derive the $\Omega_K$ if $d_P$
and $d_M$ are obtained. With the measurements of Hubble parameter,
we use GP method to reconstruct the $H(z)$ function and use equation
(\ref{dP2}) to derive the $d_P(z)$ function. Using the measurements
of angular diameter distance, transverse comoving distance can be
calculated easily through $d_M=d_A(1+z)$. We used the $H(z)$ and
$d_A$ measurements collected from previous literature. The
reconstructed $H(z)$ function and the derived $d_P(z)$ function are
shown in Figures \ref{fig2} and \ref{fig3}. In order to compare with
$\Lambda$CDM model, the best-fitted $H(z)$ and $d_P(z)$ function are
also shown in these figures. The comparison shows that the GP method
can give a reliable reconstructed function from observed data.
Figure \ref{fig4} shows the derived $\Omega_K$ at several different
redshifts, which are also listed in Table 3. Using equations
(\ref{ave_omega_K}) and (\ref{ave_sigma_omega_K}), the average
$\Omega_K$ can be obtained,  which is $\Omega_K=-0.09\pm0.19$. This
result shows that $\Omega_K$ has no significant derivation from
non-zero value.

\begin{figure}
  \includegraphics[width=\textwidth]{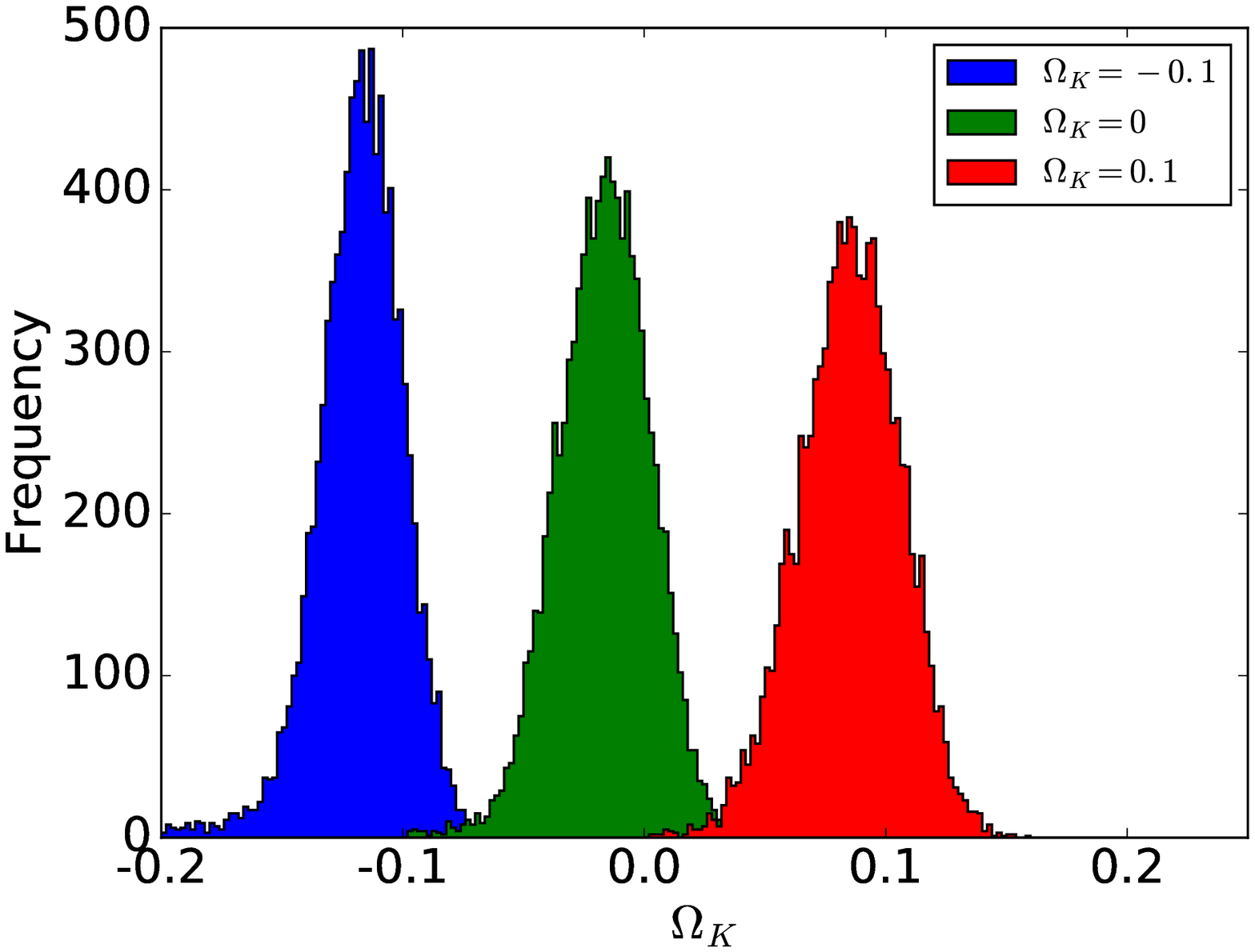}\\
  \caption{Same as Fig. \ref{fig6} but for C07 method.}\label{fig8}
\end{figure}

To check the validity and efficiency of our method, we use Monte
Carlo simulation to test it. For $\Lambda$CDM model with three
different $\Omega_K$, $-0.1$, $0$ and $0.1$, we simulate $10^4$
times for each case. Figure \ref{fig5} gives an example of the
simulations for $\Omega_K=0$ case and Figure \ref{fig6} gives the
posterior distributions of $\Omega_K$ determined with our method for
the three $\Omega_K$ cases. These two figures show that our method
can give a reliable and efficient constraint on $\Omega_K$. We also
compared our method with the DSR and C07 method. We find that DSR
method needs a higher accuracy of measurement than our method. More
importantly, the systematic uncertainty of gravitational lens system
parameter $f^2$ restricts significantly its efficiency on
constraining the curvature. For the C07 method, we also test it with
simulations. The result is shown in Figures \ref{fig7} and
\ref{fig8}. From Figure \ref{fig7}, we find that the first
derivative function of $d_M(z)$ derived from GP method has a large
derivation with the theoretical one. So the determined $\Omega_K$
not reliable. Figure \ref{fig8} shows the posterior distributions of
$\Omega_K$ determined by C07 method for the three $\Omega_K$ cases.
Meanwhile, the C07 method will give a large uncertainty on the
determined $\Omega_K$ with observed data at same accuracy level.

Future observations will improve the
constraint on the cosmic curvature. The Extended Baryon Oscillation
Spectroscopic Survey (eBOSS) will complie 250,000 new,
spectroscopically confirmed luminous red galaxies, which yield
measurements of $d_A$ with 1.2\% precision and measurements of
$H(z)$ with 2.1\% precision \citep{Dawson2016}. HETDEX will perform a
survey of 800,000 Ly$\alpha$ emission-line galaxies at $1.8<z<3.7$
\citep{Hill2006}. The precision on $d_A$ and $H(z)$ is of order 2\%
using BAO. The BAO analysis from Wide Field Infrared Survey
Telescope (WFIRST) will yield about 1.0\% measurements of the
angular diameter distance $d_A$ and Hubble parameter $H(z)$ by 17
million galaxies redshift survey in the redshift range $1.3<z<2.7$
\citep{Green2012}. Meanwhile, the Euclid satellite with survey area
of approximately 14,000 deg$^2$ and redshift range $0.7<z<2.0$ will
measure Hubble parameter $H(z)$ with 1.5\% precision
\citep{Laureijs2011}. \cite{Weinberg2013} had predicted the accuracy of future measurement of Hubble parameter and angular diameter distance through the full sky BAO survey and gave a encourage forecast that the relative error
on $H(z)$ and $d_A$ would be less than 1\% at redshift $z>0.5$.
Therefore, with our method, the curvature parameter $\Omega_K$ can
be constrained at a very high accuracy level in a model-independent
way.

\section*{Acknowledgements}
We thank the anonymous referee for useful comments and suggestions.
This work is supported by the National Basic Research Program of
China (973 Program, grant No. 2014CB845800) and the National Natural
Science Foundation of China (grants 11422325 and 11373022), the
Excellent Youth Foundation of Jiangsu Province (BK20140016).

\clearpage

\begin{table}\label{tab:d_A}
\begin{center}
\caption{The angular diameter distance $d_A(z)$ and their
references.}
\begin{tabular}{ccc}
  \hline
  $z$ & $d_A(z) \rm(Mpc)$ & Reference \\ \hline
  0.44 & $1205\pm114$ &  \\
  0.6 & $1380\pm95$ & \cite{Blake2012} \\
  0.73 & $1534\pm107$ &  \\ \hline
  0.35 & $1050\pm38$ & \cite{Xu2013} \\ \hline
  0.57 & $1380\pm23$ & \cite{Samushia2014} \\ \hline
  2.34 & $1662\pm96$ & \cite{Delubac2015} \\
  \hline
\end{tabular}
\end{center}
\end{table}

\begin{table}\label{tab:H_z}
\begin{center}
\caption{The data of Hubble parameter $H(z)$ and their references.}
\begin{tabular}{ccc}
  \hline
  $z$ & $H(z)\rm km/s/kpc$ & reference \\ \hline
  0.09 & 69$\pm$12 & \cite{Jimenez2003} \\ \hline
  0.17 & 83$\pm$8 &  \\
  0.27 & 77$\pm$14 &  \\
  0.40 & 95$\pm$17 &  \\
  0.90 & 117$\pm$23 & \cite{Simon2005}  \\
  1.30 & 168$\pm$17 &  \\
  1.43 & 177$\pm$18 &  \\
  1.53 & 140$\pm$14 &  \\
  1.75 & 202$\pm$40 &  \\ \hline
  0.24 & 79.69$\pm$3.32 & \cite{Gaztanaga2009} \\
  0.43 & 86.45$\pm$3.27 & \\ \hline
  0.48 & 97$\pm$62 & \cite{Stern2010} \\
  0.88 & 90$\pm$40 &  \\ \hline
  0.179 & 75$\pm$4 & \\
  0.199 & 75$\pm$5 & \\
  0.352 & 83$\pm$14 & \\
  0.593 & 104$\pm$13 &   \cite{Moresco2012}  \\
  0.680 & 92$\pm$8 &  \\
  0.781 & 105$\pm$12 &  \\
  0.875 & 125$\pm$17 &  \\
  1.037 & 154$\pm$20 &  \\ \hline
  0.44 & 82.6$\pm$7.8 &  \\
  0.60 & 87.9$\pm$6.1 & \cite{Blake2012} \\
  0.73 & 97.3$\pm$7.0 &  \\ \hline
  0.35 & 84.4$\pm$7.0 & \cite{Xu2013} \\ \hline
  0.07 & 69$\pm$19.6 &  \\
  0.12 & 68.6$\pm$26.2 & \cite{Zhang2014}  \\
  0.20 & 72.9$\pm$29.6 &  \\
  0.28 & 88.8$\pm$36.6 &  \\ \hline
  0.57 & 93.1$\pm$3.0 & \cite{Samushia2014} \\ \hline
  2.34 & 222$\pm$7 & \cite{Delubac2015} \\
  \hline
\end{tabular}
\end{center}
\end{table}

\begin{table}\label{tab:Omega_K}
\begin{center}
\caption{The $\Omega_K$ derived from equation (\ref{dpdm}) using our
method.}
\begin{tabular}{lccccccc}
  \hline
  $z$ & 0.35 & 0.44 & 0.57 & 0.60 & 0.73 & 2.34 & Average\\ \hline
  $\Omega_K$ & 1.24$\pm$2.78 & 0.63$\pm$3.74 & 0.29$\pm$0.75 & -0.28$\pm$1.59 & 0.13$\pm$1.18 & -0.13$\pm0.20$ & -0.09$\pm$0.19\\
  \hline
\end{tabular}
\end{center}
\end{table}

\end{document}